\documentclass[conference]{IEEEtran}

\usepackage{amsthm,amssymb,graphicx,multirow,amsmath,color,amsfonts}
\usepackage[update,prepend]{epstopdf}
\usepackage[noadjust]{cite}
\usepackage[latin1]{inputenc}
\usepackage{bbm} 
\usepackage{pdfpages}
\usepackage{tabulary}
\usepackage{multirow}
\usepackage{comment}
\usepackage{array}
\usepackage{tabularx}
\usepackage{hyperref} 
\usepackage{adjustbox}
\usepackage{tikz}
\usepackage{textcomp}
\usepackage{lipsum}
\newcommand\copyrighttext{  
	This work has been submitted to the IEEE for possible publication. 
	Copyright may be transferred without notice, after which this version 
	may no longer be accessible.
	
}
\newcommand\copyrightnotice{%
	\begin{tikzpicture}[remember picture,overlay]
	\node[anchor=north,yshift=-7pt] at (current page.north) {\fbox{\parbox{\dimexpr\textwidth-\fboxsep-\fboxrule\relax}{\copyrighttext}}};
	\end{tikzpicture}%
}

\usepackage{amsmath,amssymb}
\usepackage{caption}

\IEEEoverridecommandlockouts

\let\svbibcite\bibcite
\def\bibcite#1#2{\svbibcite{#1}{#2}}
\makeatletter
\let\svbiblabel\@biblabel
\def\@biblabel#1{\svbiblabel{#1}}
\makeatother

\begin{document}\vspace{-0.2cm}
\title{3GPP Release 18 Wake-up Receiver: Feature Overview and Evaluations
} \vspace{-0.3cm}   

\author{\IEEEauthorblockN{Andreas H\"oglund$^1$, Mohammad Mozaffari$^2$, Yanpeng Yang$^1$, Giuseppe Moschetti$^1$, \\Kittipong Kittichokechai$^1$, and Ravikiran Nory$^2$
	}\vspace{-0.05cm}\\
	\IEEEauthorblockA{
	\small  $^1$Ericsson Research, Sweden. Emails: \{andreas.hoglund, yanpeng.yang, giuseppe.a.moschetti, kittipong.kittichokechai\}@ericsson.com
 \\  
	\small  $^2$Ericsson Research, USA. Emails: \{mohammad.mozaffari, ravikiran.nory\}@ericsson.com
 \\
\vspace{-0.2cm}}
		
	}
\maketitle
\copyrightnotice
\begin{abstract}
	
	Enhancing the energy efficiency of devices stands as one of the key requirements in the fifth-generation (5G) cellular network and its evolutions toward the next generation wireless technology. Specifically, for battery-limited Internet-of-Things (IoT) devices where downlink monitoring significantly contributes to energy consumption, efficient solutions are required for power saving while addressing performance tradeoffs. In this regard, the use of a low-power wake-up receiver (WUR) and wake-up signal (WUS) is an attractive solution for reducing the energy consumption of devices without compromising the downlink latency. This paper provides an overview of the standardization study on the design of low-power WUR and WUS within Release 18 of the third-generation partnership project (3GPP). We describe design principles, receiver architectures, waveform characteristics, and device procedures upon detection of WUS. In addition, we provide representative results to show the performance of the WUR in terms of power saving, coverage, and network overhead along with highlighting design tradeoffs.

\end{abstract}

\section{Introduction}

The fifth generation (5G) of cellular communications technology and new radio (NR) developed in the third-generation partnership project (3GPP) offers increased performance and a wide range of services (e.g., ultra-reliable low-latency communications, URLLC) \cite{r1,r2}. Compared to previous generations, 5G NR has a leaner signaling design, which limits the control overhead signaling in an unloaded network thus reducing the network energy consumption. Further, 5G can operate in a wider range of frequency bands, especially with adaption to work in higher frequency bands, e.g., mm-wave bands, with a significantly reduced latency. For the device, or user equipment (UE), operation in a wider bandwidth will give a higher achievable throughput. However, this is also more energy-consuming for the UE and can lead to shorter device battery life. 

Several features have been introduced in 3GPP to reduce the device energy consumption and prolong battery life, e.g., bandwidth part (BWP) switching, monitoring of physical downlink control channel (PDCCH) in a narrower so-called control resource set (CORESET), and disabling secondary cells when not used \cite{TrPower}. In general, monitoring PDCCH is the main contributor to the UE energy consumption, and reducing the PDCCH monitoring for the UE, in the time or frequency domain, is the key factor to achieving longer device battery life. To reduce the PDCCH monitoring time, a sequence-based wake-up signal (WUS) was introduced in LTE Release 15 for low-power wide-area (LPWA) solutions narrow-band Internet of Things (NB-IoT) and long-term evolution machine type communication (LTE-M) \cite{r3}. With this solution, the UE in the radio resource control (RRC) Idle state only monitors the paging occasion (PO) if a WUS is first detected within a configured time offset before the PO. For these LPWA solutions, 20 dB coverage enhancement was introduced, mainly achieved by time repetition. The main motivation for the introduction of WUS was to reduce the monitoring time of the large number of PDCCH repetitions in the PO required for UEs in poor coverage in the case when there is no paging for the UE (which is most often the case).

In Release 16, the feature was enhanced to include UE subgroups using multiple WUS sequences to reduce the false paging, i.e., that the UE is unintentionally woken up by paging for another UE sharing the same PO. In Release 16 group WUS (GWUS), up to 8 UE subgroups per PO are supported. For NR, a downlink control information (DCI)-based WUS, referred to as DCP, was introduced in Release 16 for RRC Connected state \cite{r4}. DCP is an add-on to existing RRC Connected state discontinuous reception (DRX) where the UE will only monitor PDCCH in the configured DRX on-duration window if first a PDCCH-based WUS (DCI format 2-6) is received within a fixed time offset before the DRX on-duration. If it is not, the UE can skip the entire on-duration and will thereby reduce the energy consumption. In NR Release 17 paging early indication (PEI) was introduced. PEI is a WUS feature for RRC Idle and Inactive where, like the Release 15 WUS for LTE-M and NB-IoT, the UE only wakes up to monitor PDCCH in the PO if the PDCCH-based WUS (DCI format 2-7) associated to the PO and monitored a time offset before the PO is received first \cite{r5}. For PEI, UE subgrouping of up to 8 subgroups per PO was introduced and indicated by bits in the DCI.

For NR, the Release 16 DCP for RRC Connected and Release 17 PEI for RRC Idle/Inactive were mainly intended for mobile broadband (MBB) use cases and human-originated traffic. For WUS in general, the biggest gain can be achieved when relatively low downlink latency needs to be achieved while there is rarely anything to transmit to the UE. That is, if battery life is the only relevant performance metric, the UE can spend most of the time in a sleep state, e.g., using the existing extended DRX (eDRX) or Mobile Initiated Communication Only (MICO) features, yet, the drawback is that the downlink latency can be several hours long. Therefore, the benefit of WUS is that the UE power saving can be achieved without compromising the latency. 

The UE power saving gain is primarily achieved by keeping the main receiver (MR) in a sleep state to conserve energy. If the MR does not have to be started every time the UE monitors WUS but can be kept in a sleep state, large UE power saving can be achieved compared to existing DRX/eDRX. Gains will therefore be bigger for more infrequent data transmissions, or less active traffic models, which is typically the case for IoT services, as opposed to the previous NR DCP and PEI features which are more focused on more active MBB use cases. To this end, a separate wake-up receiver (WUR) must be used, and this is the new aspect studied by 3GPP in Release 18 \cite{r6}. While a similar WUR solution has been studied and introduced in IEEE for WiFi \cite{r7}, the 3GPP solution is different in terms of design and requirements. The new aspects with the Release 18 WUR include the new type of receiver required, which is covered in Section II-A, the design of a WUS suitable for this receiver, covered in Section II-B, and if any changes to procedures are required, covered in Section II-C. Evaluation results are then presented in Section III to see how the new solution performs compared to prior solutions. Finally, Section IV covers a summary and conclusions from the work.

 \section{Low Power Wake-up Receiver And Signal}
In this section, we present different WUR architectures, signal design, waveform generation, and WUR monitoring procedures. 
 \subsection{WUR Architecture}
Among different receiver architectures, two common WUR architectures are: the direct demodulation approach, also known as radio frequency (RF) envelope detector (ED), and the on-chip local oscillator (LO) approach. The first type of architecture is characterized by low complexity, low cost, and extremely low energy consumption since no active RF circuits are typically used. In contrast, the second type of architecture requires more complex components, like on-chip local oscillators that down-convert the incoming RF signal to baseband (BB) or intermediate frequency (IF). This results in higher energy consumption relative to the ED, but the architecture can offer better sensitivity and robustness to interferers thanks to the possibility of implementing sharp BB/IF filters. In the on-chip LO Zero-IF architecture, also known as a homodyne receiver, the incoming RF signal is directly down-converted to BB. A schematic of on-chip LO Zero-IF is shown in Figure 1a. The RF input signal is first selected by a band-pass filter (BPF) and then amplified by an RF low-noise amplifier (LNA). Typically, a matching network is present between the BPF and the RF LNA to easily provide the optimum noise impedance at the input of the LNA. The RF signal is then directly down-converted to BB by a mixer and further amplified by a low-power BB amplifier. To minimize energy consumption, the local oscillator is not locked in a phase-locked loop (PLL). The BB signal is hence filtered by a low-pass filter (LPF) and converted to the digital domain using an analog-to-digital converter (ADC). Thanks to its simplicity, this architecture allows a high level of integration, but the direct conversion can cause problems such as direct current (DC) offset due to LO leakage, second-order inter-modulation products, and large flicker (1/f) noise. The energy consumption of this architecture could be further reduced by using a mixer-first approach, removing the RF LNA, at the penalty of higher noise figure.

Another candidate being discussed in 3GPP Release 18 is the orthogonal frequency division multiplexing (OFDM) WUR architecture. Even though an OFDM-capable architecture is more complex and more power-hungry than the simpler Zero IF architecture described above, it can bring some advantages. For instance, such WUR would be capable of in-phase and quadrature (I/Q) sampling and measurements from existing NR signals, thus allowing less frequent wake-up of the MR, which significantly can reduce the energy consumption. A proposed architecture for an OFDM-capable WUR is shown in Figure 1b. In this architecture, similarly to the zero-IF WUR, a BPF, a matching network, and an RF LNA are present at the input, after the receiving antenna. After the RF LNA, the signal is down-converted to BB by an I/Q mixer driven by LO with PLL. The signal is then amplified by a BB amplifier, filtered by LPFs and converted to a digital domain by ADCs. Depending on the implementation, the expected energy consumption of the OFDM WUR is around 10-20 times lower than the MR while it can be roughly 3-5 times higher than the OOK WUR \cite{r6}. 

    \begin{figure}[!t]
	\begin{center}
		\includegraphics[width=8.5cm]{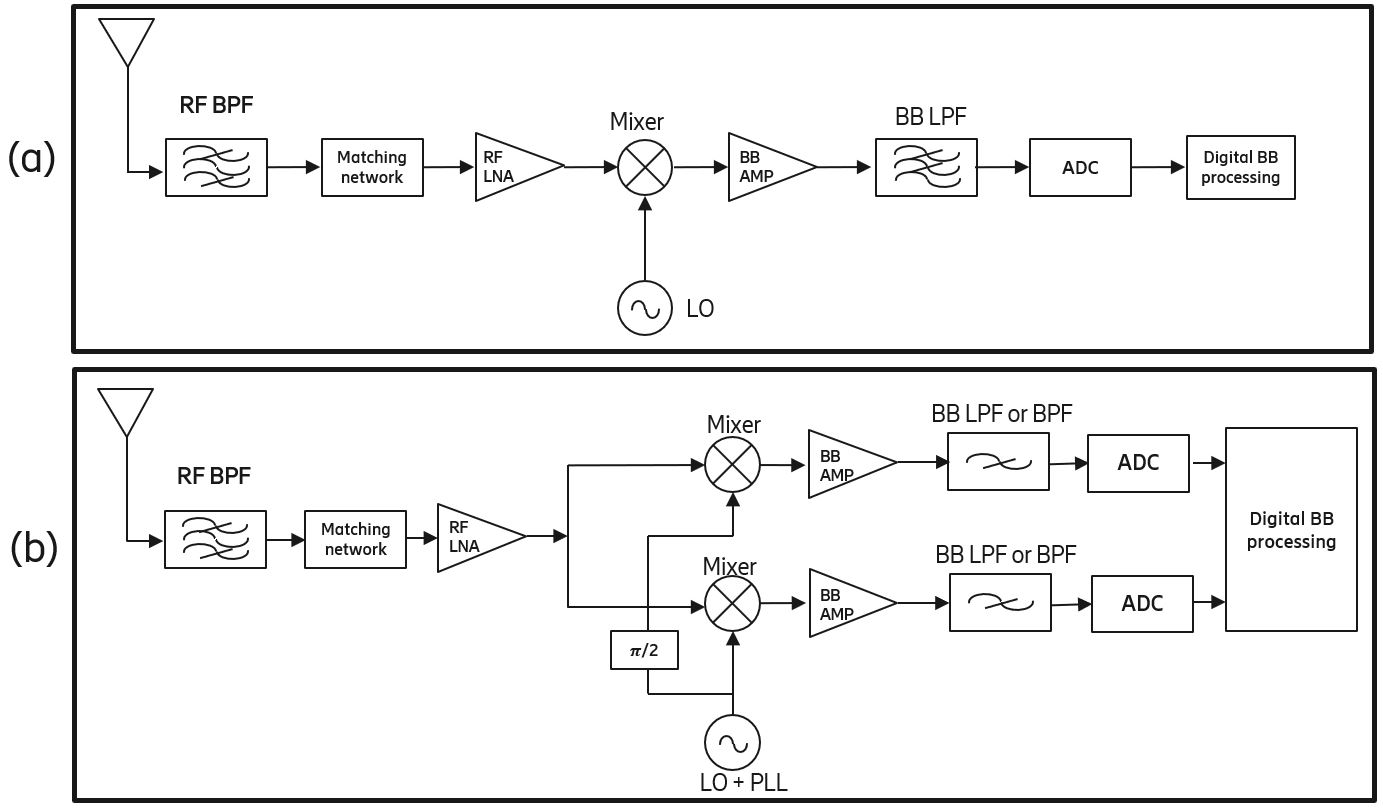}
		\caption{(a) Schematic of on-chip LO Zero-IF architecture. (b) Schematic of OFDM architecture.}\vspace{-0.02cm}
		\label{receiver}
	\end{center}
\end{figure}  

\subsection{Wake-up Signal Design}
To ensure that the WUR can detect the WUS and perform necessary functionalities, the signal design needs to consider the receiver architecture, performance requirements as well as network and coexistence impacts. The key design principles are as follows: 1) it should be possible to multiplex the WUS with other NR transmissions in time or frequency domain without causing interference and 2) it should be possible to generate the WUS with the existing hardware of the 5G base station (gNB) transmitter without creating new emissions/compliance requirements. The signal design involves determining a suitable modulation and coding scheme, signal structure, payload, and time-frequency span of the signal. 
Regarding the WUS bandwidth, the 3GPP recommended a bandwidth less than or equal to 5 MHz for Idle/Inactive mode although other bandwidth sizes up to 20 MHz can be considered \cite{r6}. Also, for multiplexing with other NR signals and channels, it is beneficial if WUS has a flexible frequency position such that it can be flexibly allocated within a carrier.  During the 3GPP Release 18 study, LP WUS performance evaluations were conducted for different payload sizes, e.g., from 1 to 48 bits, to carry various contents such as UE group/sub-group ID or UE ID. While there was no conclusion on the payload size, it is expected that WUS can carry a small to medium payload (e.g., 8 bits) to provide a similar subgrouping benefit as in 3GPP Release 17 PEI. 

In terms of modulation, the two main candidates for WUS in 3GPP Release 18 are OOK-based WUS and OFDM-based WUS, each with their pros and cons. 

\subsubsection{OFDM-based WUS}
 The existing OFDM-based NR signal structure can be reused for transmitting WUS with a minimum or no impact on the base station (i.e., gNB transmitter) for the waveform generation. For example, the existing 5G Secondary Synchronization Signal (SSS) and other reference signals along with existing sequences, such as m-sequence and Zadoff-Chu sequence, can be reused for transmitting WUS. Such sequences have \emph{good} autocorrelation and cross-correlation properties, making it possible to perform correlation-based detection both in the time and frequency domain with a desirable detection performance. The receiver architecture corresponding to OFDM-based WUS (i.e., OFDM WUR as described in the previous section) needs to be able to process  I/Q samples and extract phase information of the received signal, leading to additional capabilities and enhanced detection performance \cite{r6}. The OFDM-based WUR/WUS has the following benefits: 1) compared to OOK-based WUS, the OFDM waveform can reach a target coverage with lower resource consumption, 2) the WUR capable of receiving OFDM waveform can reuse existing NR synchronization signals to perform radio resource management (RRM) measurement and synchronization, thus avoiding the introduction of a new always-on synchronization signal, and 3) minimum impact on the base station for generating WUS in coexistence with other NR transmissions.

\subsubsection{OOK-based WUS}
\begin{figure}[!t]
	\begin{center}
		\includegraphics[width=8.9cm]{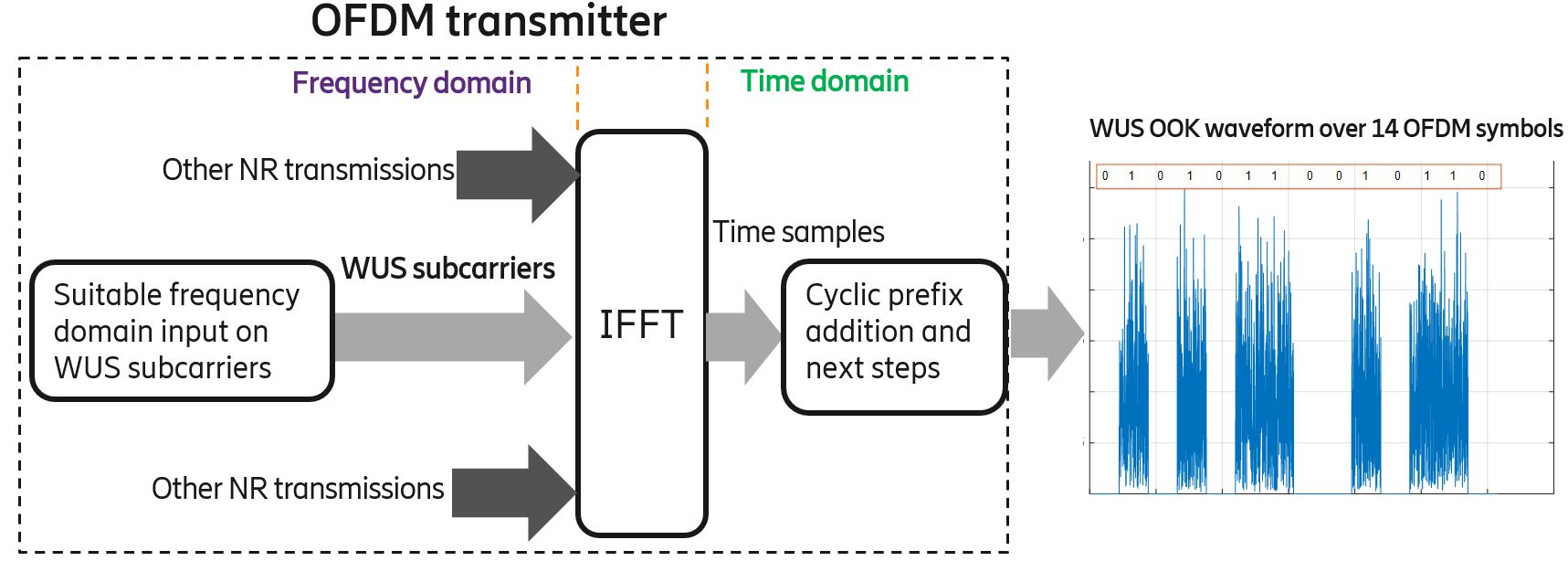}
		\caption{Example of generating an OOK waveform with an OFDM transmitter.}\vspace{-0.02cm}
		\label{Signal}
	\end{center}
\end{figure}

On-off keying (OOK) waveform is a special form of amplitude-shift keying where the information is carried through a sequence of ON (i.e., high power level) and OFF (i.e., low power level) signals. An OOK waveform is attractive for low power and low complexity receivers as it can be detected with an envelope detector in the time domain without the need for power-hungry components such as an accurate oscillator and PLL. While the OOK WUS provides power saving benefits for the WUR, it has less coverage compared to OFDM-based signals for the same resource overhead. Consequently, to reach a target coverage, the system overhead in terms of time-frequency resource consumption is higher than for an OFDM-based WUS. Another consideration is that the OOK waveform needs to be generated using the existing base stations while ensuring efficient coexistence with existing OFDM-based transmissions. The two main variants of the OOK waveform generated by an OFDM transmitter are as follows.
\begin{itemize}
    \item  Single-bit OOK: within one OFDM symbol, only one ON/OFF OOK segment is transmitted. OOK WUS can be generated by transmitting one bit (0 or 1) per OFDM symbol. In this case, to generate ``1" WUS subcarriers have non-zero power (e.g., random QAM symbols) while ``0" is generated by having zero-power WUS subcarriers. Single-bit OOK generation of ON/OFF signal is straightforward with minimum impact on the OFDM transmitter.

    \item Multi-bit OOK: to increase the data rate, multiple ON/OFF OOK segments are transmitted within one OFDM symbol.  Specifically, the OFDM transmitter should generate a time domain signal that is close to a desired OOK waveform.  To ensure a minimum impact on the transmitter and avoid inter-subcarrier interference, the inputs to the Inverse Fast Fourier Transform (IFFT) in the frequency domain need to be determined such that the output of the IFFT (in the time domain) represents a desired time domain signal \cite{TicLetter}. Compared to single-bit OOK, the generation of the multi-bit OOK waveform is not as straightforward at the base station. Nonetheless, the complexity of waveform generation can be reduced by pre-storing the generated frequency domain samples which are mapped to the WUS sub-carrier segment of IFFT at the base station. Additionally, mapping the generated frequency domain values (before IFFT) to existing sequences/QAM modulations is beneficial for implementation and it reduces the impact on the base station \cite{ContributionSignal}. Figure \ref{Signal} shows an example of generating an OOK waveform with an OFDM transmitter. 
\end{itemize}

In addition, a  harmonized design based on both OOK and OFDM WUS can be considered where the signal can be received by OOK-based WUR and OFDM-based WUR \cite{WURWID}. In this case, OFDM sequences can be additionally modulated on top of the OOK waveform to carry information and provide benefits for devices supporting OFDM-based WUR.

\subsection{Procedures Mobility Measurements}
\begin{figure}[!t]
	\begin{center}
		\includegraphics[width=8.7cm]{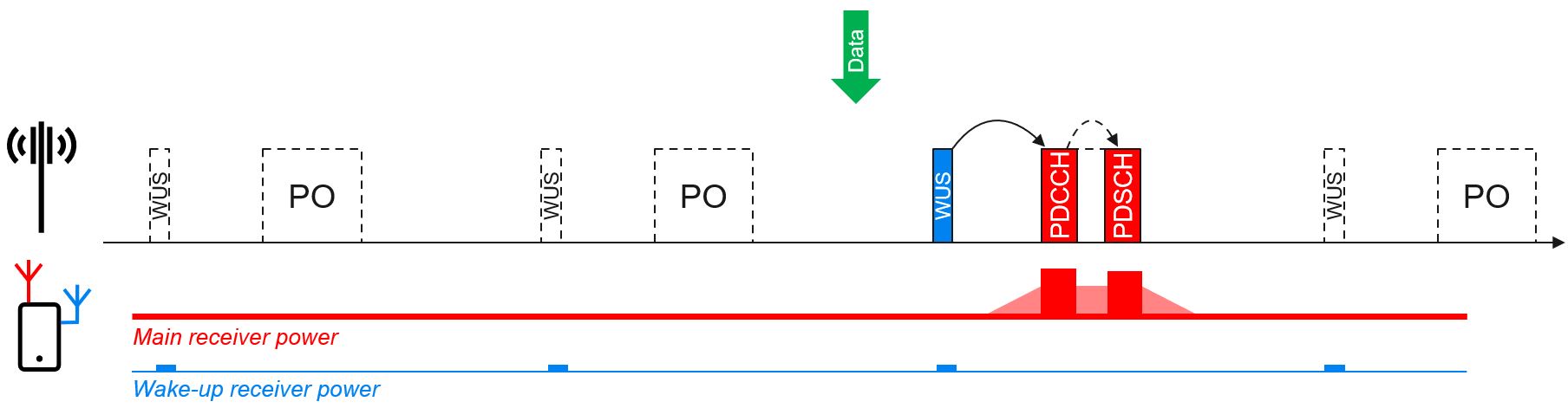}
		\caption{Illustration of WUR monitoring procedure and UE power consumption.}\vspace{-0.02cm}
		\label{Procedure}
	\end{center}
\end{figure}

The WUR power saving benefit comes from keeping the MR in a sleep state and turning off the most power-hungry components in the UE. Therefore, a time offset or gap between the UE detection of the WUS and the resumption of UE procedures is required to cater to the start-up time of the main receiver. 
In Idle/Inactive mode, this would typically be before the PO in which the UE monitors downlink control information on PDCCH to see if there is incoming data for the UE. An illustration of WUR monitoring procedure is presented in Figure \ref{Procedure}.

In the Connected mode, the WUS monitoring occasion could be before the DRX on-duration in which the UE monitors PDCCH, or a new separate PDCCH monitoring could be defined for WUR operation. The difference between the two is minor but in general, means a different duty or DRX cycle length and PDCCH monitoring window can be applied for WUR operation, i.e. not dictated by legacy DRX configuration. A short WUR duty-cycle length is beneficial for WUR since it reduces the downlink latency while UE energy consumption can still be kept very low since MR is in a sleep state. If the WUS monitoring occasions are not tied to a legacy procedure, the WUR duty-cycle could be configured freely or WUR operation could even be continuous in time. In this case, the WUR is not switching between active and sleep modes but constantly remains in the active mode. Therefore, due to the higher rate of WUS false alarms (see Section III-B), the continuous WUR operation will have a higher energy consumption than duty-cycled WUR operation. The potential benefit of continuous WUR is shorter downlink latency, but this is only the case in practice if the procedure triggered by WUS detection is not restricted to a certain periodicity and hence determines the latency. For example, in Idle the latency will be dictated by the periodicity of the physical random access resources.

\section{Performance Evaluation Results}

Here, we evaluate the UE power saving gain obtained by WUR assuming ultra-deep sleep for the MR in RRC Idle state. Besides power saving on the UE side, sending LP-WUS will consume extra radio resources and energy for the network. To this end, the network impact in terms of resource overhead and network energy consumption is also evaluated. Another aspect to consider is WUR coverage as the WUR is less capable than the MR. To ensure the same service, the WUR would need to have at least the same coverage as the weakest existing NR physical channel. 

\subsection{Coverage}
\begin{figure}[!t]
	\begin{center}
		\includegraphics[width=8cm]{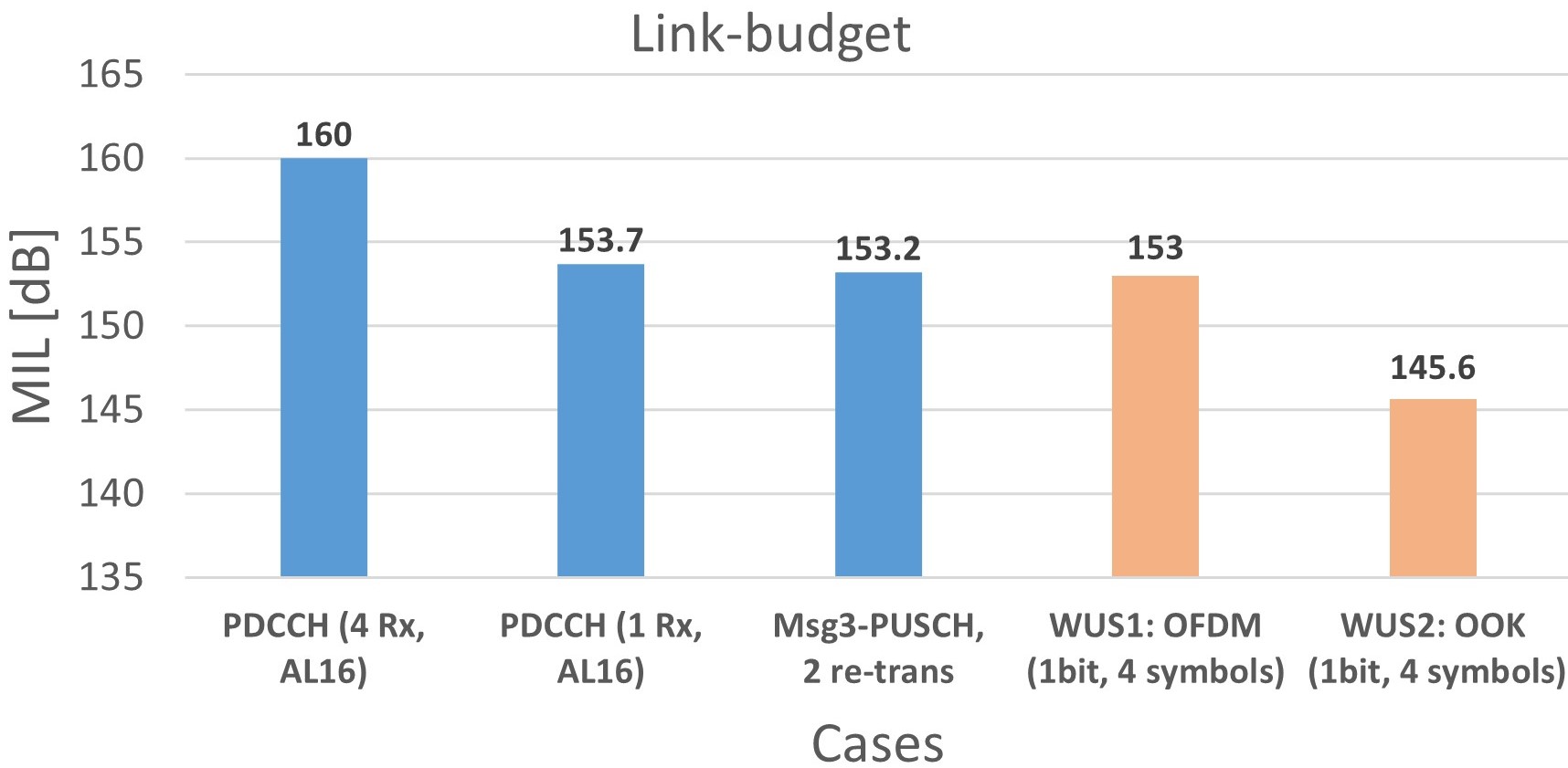}
		\caption{Link-budget performance of WUS and reference channels.}\vspace{-0.02cm}
		\label{Coverage}
	\end{center}
\end{figure}

The coverage of WUR depends on several aspects such as WUR modulation, WUS payload, duration, the bandwidth of the WUS, and the WUR architecture. To determine the coverage, both link-level and link-budget evaluations need to be performed. Here, based on the 3GPP coverage evaluation methodology \cite{r6}, we provide coverage results for different WUR alternatives and compare them with two NR channels (as reference): PDCCH for paging and Msg3 physical uplink shared channel (PUSCH), which are used during the initial/random access procedure. The main assumptions for the evaluations are: 2.6 GHz carrier frequency for urban environment, TDL-C channel model, 300 ns delay spread, 3 km/h UE speed, 5 MHz WUS bandwidth, number of WUR antennas is 1 (1 Rx), and the coverage metric for link-budget analysis is the maximum isotropic loss (MIL). Further details on the 3GPP evaluation assumptions and link-budget template are available in \cite{r6}. We consider the following cases for performance comparison in Figure \ref{Coverage}:
\begin{itemize}

  \item PDCCH detected by the main receiver:  aggregation level (AL) 16, number of receiver branches (Rx): 1 or 4.
  \item MSG3-PUSCH detected by the main receiver: two hybrid automatic repeat request (HARQ) retransmissions.
  \item WUS1: OFDM-based WUS detected by OFDM WUR with a noise figure 3 dB worse than the main receiver. WUS duration for sending 1 bit is 4 OFDM symbols.
  \item WUS2: OOK-based WUS detected by OOK WUR with a noise figure 6 dB worse than the main receiver. WUS duration for sending 1 bit is 4 OFDM symbols.
\end{itemize}

As we can see in Figure \ref{Coverage}, the coverage of OFDM WUS is at least 7 dB better than that of OOK WUS. This is because of the enhanced detection performance of the OFDM WUR and its lower noise figure compared to the OOK WUR. The performance of the considered OFDM WUS is close to Msg3 and PDCCH with 1 Rx (e.g., RedCap devices \cite{RedCapMags}). However, to match the coverage of PDCCH with 4 Rx, the WUS duration needs to be increased. It is also observed that it is challenging for the OOK WUS to meet the coverage of reference channels, especially the PDCCH. While the coverage can be improved by increasing the WUS duration, it results in a significant network overhead for the OOK WUS, particularly for large payload sizes.  Therefore, to achieve full coverage in the cell, the OOK WUS should only carry a small payload (up to a few bits).

\subsection{UE Energy Consumption Reduction}

The purpose of using WUR is to reduce UE energy consumption by allowing the MR to remain in a sleep mode for a longer period of time. Until 3GPP Release 18 the lowest MR sleep state considered for NR was ``deep sleep" which consumes 1 unit of power and with an associated 20 ms time and 450 energy units for state transitions \cite{TrPower}. In the 3GPP Release 18, a new ``ultra-deep sleep" state for MR was introduced with 0.015 power units, but at the cost of longer transition time (400 ms) and higher transition energy (15000 energy units). Note that in the 3GPP power saving evaluation methodology, 1 unit of power is a relative power for a regular 5G device in the deep sleep mode \cite{TrPower}.

The UE energy consumption depends on MR and WUR energy consumption as well as the MR wake-up rate. For MR energy consumption, we assume the UE is of RedCap type \cite{RedCapMags} and follow the assumptions in \cite{38875}. Regarding WUR active power, we assume power values smaller than 0.5 power unit for continuous monitoring of OOK-based WUR and different values from 0.5 to 10 power units for discontinuous monitoring, which covers both OOK and OFDM-based WUR \cite{r6}.  

The UE wake-up rate is determined by the following aspects: 

\begin{itemize}
  \item UE paging rate: detection of LP-WUS addressed for the UE itself, depends on traffic model.
  \item False paging: detection of LP-WUS intended for other UEs in the same group, depends on the number of UEs per group.
  \item False alarm rate: false detection of LP-WUS due to noise/interference, depends on receiver design.
\end{itemize}

\begin{figure}[!t]
	\begin{center}
		\includegraphics[width=8.5cm]{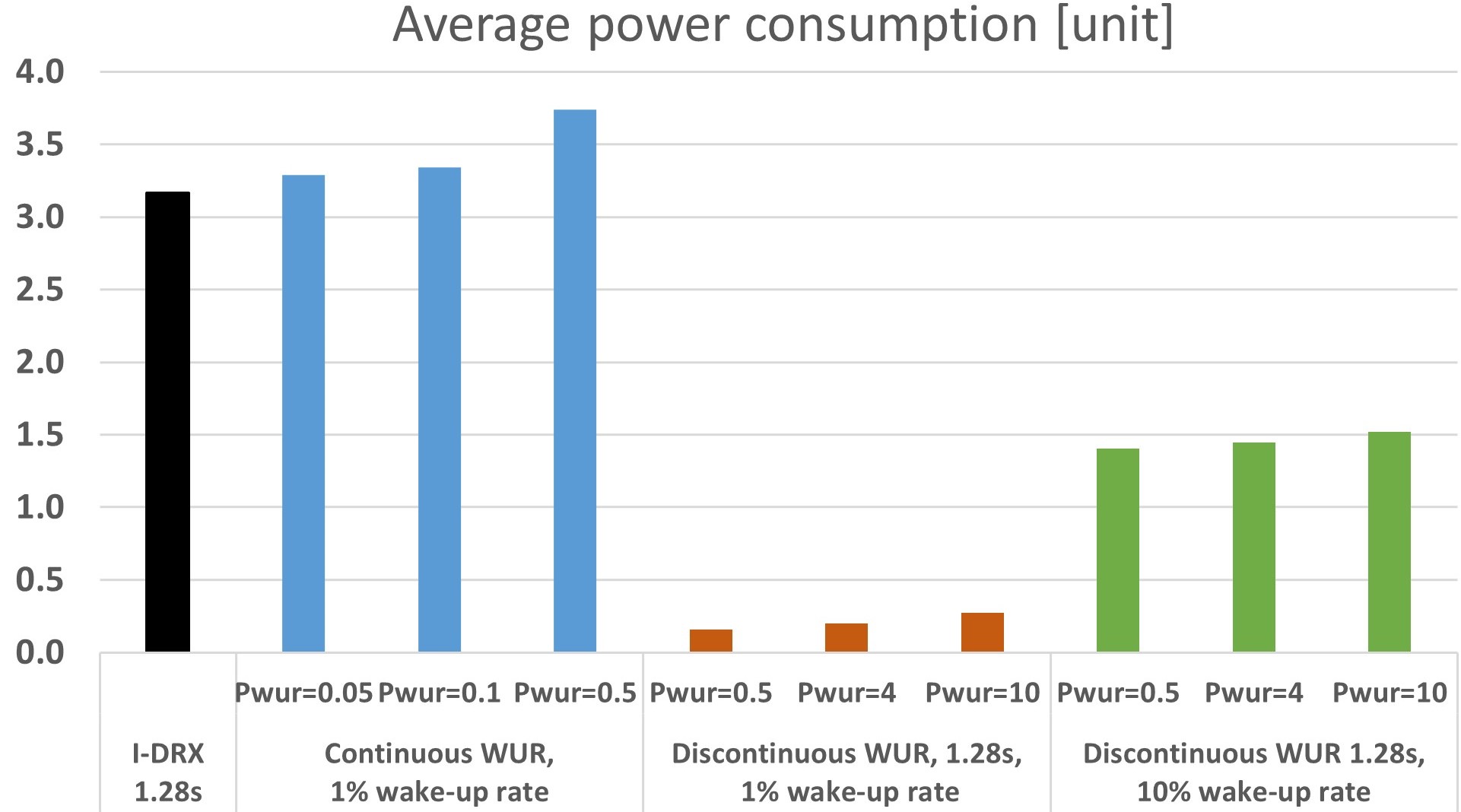}
		\caption{UE energy consumption for different cases.}\vspace{-0.02cm}
		\label{Power}
	\end{center}
\end{figure}

Figure \ref{Power} depicts the energy consumption of both discontinuous and continuous WUS monitoring with various WUR power (Pwur) and wake-up rate for a UE in RRC Idle state. As shown in the figure, discontinuous monitoring could enable over 90\% UE power saving gain compared to legacy Idle DRX because MR can stay much longer in a sleep state. Whereas with continuous WUR, the UE consumes even more energy than the baseline due to higher chance of WUS false alarms which leads to the MR waking up unnecessarily. Another observation is that a higher wake-up rate causes a dramatic increase in energy consumption due to the high transition energy. However, for discontinuous monitoring, WUR active power itself has a marginal impact on the overall energy consumption as the ON duration of WUR is quite short compared with the duty cycle length. This indicates that a more capable receiver (e.g., OFDM-based) could be a good candidate for WUR, especially when operating with a duty-cycle.

\subsection{Signaling Overhead and Network Energy Consumption}

The overhead of LP-WUS/WUR operation depends on resources needed for WUS transmission including any guard bands and resources for the WUR synchronization signal. The total overhead then depends on the traffic and typically increases with the number of WUS transmissions needed (e.g., corresponding to the paging rate). It also depends on the WUS missed detection rate as more WUS resources will be used to eventually wake up the UE.  The overhead of a single LP-WUS transmission can be determined as the fraction of the number of resource elements (REs) used for LP-WUS (including guardband and resources used for synchronization) and the total number of available REs for downlink transmission. 

In Table \ref{Overhead}, we evaluate the overhead in Idle state assuming 250 UEs per cell and 4 UEs in the same paging subgroup with per-UE paging rate of 0.1\% (further details are provided in \cite{ContributionEval}). We consider different examples of resources used for LP-WUS required to match the paging PDCCH link budget (2-OFDM symbol with AL16) with 1 or 2 Rx and Msg 3 PUSCH with 0 or 2 retransmissions, assuming 6 dB and 3 dB worse noise figures for the OOK-based and OFDM-based WUS receivers, respectively, when compared to that of the OFDM-based receiver with 1 Rx antenna for PDCCH. For the OOK WUR, a new low-power synchronization signal (LP-SS) is also considered with the overhead of 3 slots per 1.28\,s. From the table, the overhead for OOK-based WUS (e.g., Pwur=0.5) is roughly ~50-100x higher compared to that of OFDM-based WUS (e.g., Pwur=4) as the OOK-based WUS duration required to achieve the same coverage target as OFDM-based WUS is much higher and there is a need for additional periodic LP-SS transmissions.

\begin{table}[t]
	\centering 
\caption{\small System overhead of LP-SS and LP-WUS.}
\label{Overhead}
\begin{adjustbox}{width=\columnwidth,center}

	\begin{tabular}{|c|c|c|c|c|} \hline 
	&	Msg3 0rtx &	Msg3 2rtx 	& PDCCH 1Rx	&PDCCH 2Rx           \\\hline
		\shortstack {\\OOK WUR \\(Pwur = 0.5)} &	0.46\%	& 0.90\% &	1.15\% &	2.37\% \\ \hline 
		\shortstack {\\OFDM WUR\\ (Pwur = 4)} &	0.016\%	& 0.016\%	&0.016\% &	0.02\%
		\\ \hline                          
	\end{tabular}\vspace{0.2cm}
 \end{adjustbox}
\end{table}

Regarding network energy consumption, we focus on the extra consumption due to additional transmission of the WUS synchronization signal, i.e., the LP-SS. Therefore, the energy consumption of a network with zero load is our baseline, i.e., only periodic \emph{always on} transmissions, SSB, LP-SS, and SIB1, are considered. For OOK-based WUR, we assume that LP-SS occupies 12 PRBs in the frequency domain and 4 to 42 symbols per beam in the time domain with 8 beams in total. The periodicity varies from 320 ms to 10240 ms. For frequent LP-SS transmissions (e.g., LP-SS used for RRM with a periodicity of 320 ms), it results in an additional network energy consumption of up to 11\%. For infrequent LP-SS transmissions (e.g., LP-SS not used for RRM but only used as a timing reference for LP-WUR monitoring window determination with periodicity of 2560 ms), it results in an additional network energy consumption of up to 1.5\%. Therefore, the additional energy consumption for LP-SS transmission compared to baseline is not marginal unless it is restricted to only a short duration in the time domain and infrequent transmissions. However, for OFDM-based WUR, the legacy PSS/SSS in SSB can be reused for synchronization, which does not increase the network energy consumption.


%

\section{Conclusions}\vspace{-0.01cm}
In this paper, we presented an overview of 3GPP Release 18 standardization study on the design of low-power WUR and WUS. In addition to describing WUR architectures, signal design, procedures, and waveform generation schemes for  OOK-based and OFDM-based WUS, we provided various performance evaluation results. It was shown that the WUR can provide around 90\% power saving gain in RRC Idle mode. In terms of coverage, the OFDM WUS significantly outperforms the OOK WUS and can reach a coverage target (Msg3 or PDCCH) with a lower overhead (e.g., by a factor of 50). The OOK WUR has a lower active energy consumption and lower complexity compared to the OFDM WUR, but its active duration needs to be longer than the OFDM WUR which may result in a higher average energy consumption. Meanwhile, as the OFDM-based WUR can reuse the existing NR signals for various purposes, it can reduce the network overhead and energy consumption, as well as minimize the gNB complexity for transmitting WUS. In 3GPP Release 19 \cite{WURWID}, the details of signal design, WUR procedure, synchronization, and measurements will be specified.

\def\baselinestretch{1.00}
\bibliographystyle{IEEEtran}
\bibliography{referenceConf}
\vspace{0.9cm}

\begin {comment}

\textbf{Sandeep Narayanan Kadan Veedu} (sandeep.narayanan.kadan.veedu@ericsson.com) is a Senior Researcher with Ericsson Research, Stockholm, Sweden, where he is involved in the research and standardization of cellular IoT technologies. He is also a standardization delegate for Ericsson in 3GPP. Prior to joining Ericsson in 2018, he was a Research Associate with King's College London, U.K. He has also held various research positions with University College Dublin, Ireland, Northwestern University, USA, and The University of Edinburgh, U.K. He has a Ph.D. degree in electrical and information engineering from the University of L'Aquila, Italy.

\vspace{0.1cm}

\textbf{Mohammad Mozaffari} (mohammad.mozaffari@ericsson.com) is a senior researcher at Ericsson Research, Silicon Valley, USA. He received the Ph.D. degree in electrical and computer engineering from Virginia Tech in 2018. His research interests span diverse areas, including 5G and 6G wireless networks, UAV and drone communications, IoT, and machine learning. He has been actively contributing to 3GPP standardization for NB-IoT, LTE-MTC, RedCap, and wake-up radio.  He is the recipient of the IEEE ComSoc Young Author Best Paper Award and Outstanding Ph.D. Dissertation Award. He is currently an associate editor of the IEEE Vehicular Technology Magazine. \vspace{0.05cm}

\textbf{Andreas Höglund} (andreas.hoglund@ericsson.com) has a Master of Science in engineering physics (2002) and a Ph.D. in condensed matter physics from Uppsala University (2007). He is a master researcher at Ericsson Research and has since he joined in 2008 worked on HSPA, LTE, system simulations, 5G research in the METIS project, 3GPP standardization of LTE MTC (Cat-M1) and NB-IoT. He is currently working with 3GPP Rel-17 RedCap and Small Data Transmission (SDT), Rel-18 Wake-up receiver, eRedCap, MT-SDT, as well as IoT research for 6G.
\vspace{0.1cm}

\textbf{Emre A. Yavuz} (emre.yavuz@ericsson.com) received his B.Sc. and M.Sc. degrees in Electrical \& Electronics Engineering at METU in Turkey. He received his Ph.D. degree in Electrical \& Computer Engineering at UBC, Canada. He worked as a consultant in Vancouver prior to joining School of Electrical Engineering at KTH, as a post-doctoral fellow. He is a Master Researcher at Ericsson Business Unit Networks, working with 4G/5G Layer2/3 standardization in 3GPP TSG RAN WG2 leading the technical work to standardize features such as LTE MTC and NB-IoT in Rel 13-17, NR RedCap, and NR and IoT NTN  in Rel 17. 
\vspace{0.1cm}

\textbf{Tuomas Tirronen} (tuomas.tirronen@ericsson.com) is a Master Researcher at Ericsson Research, which he joined in 2012. He received his M.Sc. in telecommunications in 2006 from Helsinki University of Technology and D.Sc. (Tech.) in communications engineering in 2010 from Aalto University, Finland. He has been working as researcher in areas of IoT and machine-type communications in various projects and has been working in 3GPP standardization of LTE-M, NB-IoT, NR RedCap, and other features. He is currently vice head of Ericsson 3GPP RAN2 delegation. 
\vspace{0.1cm}

\textbf{Johan Bergman} (johan.bergman@ericsson.com) is a Master Researcher at Ericsson Business Unit Networks. He received his Master's degree in Engineering Physics from Chalmers University of Technology in Sweden. He joined Ericsson in 1997, and since 2005 he has been working with 3G/4G/5G physical layer standardization in 3GPP TSG RAN Working Group 1. As Rapporteur of the 3GPP TSG RAN Work Items on LTE-MTC in Releases 13-16, and NR RedCap in Release 17 he has led the technical work to standardize new features dedicated to IoT. He was a corecipient of Ericsson's Inventor of the Year award for 2017.
\vspace{0.1cm}

\textbf{Y.-P. Eric Wang} (eric.yp.wang@ericsson.com) is a Research Leader at Ericsson Research. He holds a PhD degree in electrical engineering from the University of Michigan, Ann Arbor. In 2001 and 2002, he was a member of the executive committee of the IEEE Vehicular Technology Society. He was an Associate Editor of the IEEE Transactions on Vehicular Technology from 2003 to 2007. He is a technical leader in Ericsson Research in the area of IoT connectivity. Dr. Wang was a corecipient of Ericsson's Inventors of the Year award in 2006 and contributed to over 200 US patents and 50 IEEE articles.

\end{comment}

\end{document}